\documentclass[10pt,a4paper,twoside]{article}
\usepackage{epsfig}
\usepackage{baltlat6}
\usepackage{array}
\usepackage{here}
\begin{document}
\ \
\vspace{0.5mm}
\setcounter{page}{277}

\titlehead{Baltic Astronomy, vol.\,??, ??--??, 2015}

\titleb{DETAILS OF THE SPATIAL STRUCTURE AND KINEMATICS\\
OF THE CASTOR AND URSA MAJOR STREAMS}

\begin{authorl}
\authorb{S.V. Vereshchagin}{1},
\authorb{N.V. Chupina}{1}
\end{authorl}

\begin{addressl}
\addressb{1}{Institute of Astronomy of the Russian Academy of Sciences (INASAN),\\
48 Pyatnitskaya st., Moscow, Russia;
svvs@ya.ru}
\end{addressl}

\submitb{Received: 2015 July 2; accepted: 2015 December 15}

\begin{summary} 
A list of the Castor stream members is compiled based on the data from various authors. 
The membership probabilities for some stars are revised based on the individual apex, multiplicity,
observational errors, and peculiarity. The apex of the Castor moving group is determined using
the apex diagram method. 
The parameters of the Castor and Ursa Major streams are compared and 
the positions of the two streams on the apex diagram are found to differ by $225^\circ$,
implying that  the two groups move in almost opposite directions. 
Stars of both moving groups are intermixed in space, the Castor stream occupies a smaller volume
than UMa stream and is located inside it. 
Our results can be useful for understanding the morphology of the Galactic disk in the Sun's vicinity.
\end{summary}


\resthead{Castor and Ursa Major streams}
{S.V. Vereshchagin, N.V. Chupina}

\sectionb{1}{INTRODUCTION}

Our analysis of stellar groups is based on the use of their proper motions, radial velocities, and 
parallaxes. The accuracy and reliability of these parameters depend
on the distance to the stars. 
To obtain the most reliable results, here we investigate the Castor and Ursa Major (UMa) streams,
which are the closest to the Sun and therefore may provide insight into 
the structure of the Galactic disk in the solar vicinity.

A feature of both streams is that they have many multiple systems, 
some of which consist of up to six stars! 
For example, the UMa stream contains a sextuple system, which includes Mizar and Alcor (Mamajek et al. 2010) 
and, perhaps HD 76644 (Zhuchkov et al. 2006). 
Moreover, UMa stream contains three kinematic groups. 

To investigate the space motions of the stars, 
we use the AD-method that we developed earlier. 
This method consists in searching for regularities in the 
positions of "individual star apexes". 
The individual star apex is the point in the sky 
indicating the direction of the space velocity of the star 
if the velocity vector starts from the origin of the equatorial system. 
The right ascension and declination of this point are denoted as $A$ and $D$,
respectively. 
This method is good at showing details of motions and 
can be used to reliably detect the kinematic structure.

The layout of this paper is as follows. We first compile the list of the Castor stream
objects with the membership revised using the AD-method and stars with large errors 
discarded. We then compute the apex of the Castor stream and compare
its parameters  with those of the UMa stream.
We find the apex positions and the structure of the AD-diagram to  differ for the two streams. 
We also compare the positions of the streams in space and discuss the results.

\sectionb{2}{BRIEF DATA DESCRIPTION}

We use the results of five studies. 
Caballero (2010)  the candidate objects of the Castor moving group (70 stars) lists in his Table~1. 
Anosova \& Orlov (1991) identified 10 stars 
as probable members this group. 
Agekyan \& Orlov (1984) found nine moving groups in the solar vicinity, 
the space motion of their group V is close that of the Castor group. 
Barrado and Navascues (1998) used the Hipparcos catalog (hereafter referred to as HIP) data
to identify 16 stars as bona fide members of the Castor stream. 
Shkolnik et. al. (2012)  found six additional candidates from their study of M-type
dwarfs. 

We use the Caballero (2010) list, the most extensive of the five, as a basis for our
compilation. We add to it three stars from Anosova \& Orlov (1991), seven stars
from Agekyan \& Orlov (1984), and six stars from Shkolnik et al. (2012). 
We thus have a total of 86 stars (all stars identified by Barrado y Navascues (1998) are already
included in the list of Caballero (2010)). 
We adopted the radial velocities for stars of our combined list from
SIMBAD database  and the remaining parameters from HIP (for stars included into this catalog). 
For the six stars from Shkolnik et al. (2012) we adopt all their parameters from that
paper.  In the next section we prepare the list of Castor moving group stars, 
determine their membership using the AD diagram, and
discard stars with large errors, as well as the runaway star. 
We will publish the list of stars of Castor moving group stars later.

\sectionb{3}{THE CASTOR AD-DIAGRAM}

Only 70 stars in our combined list have a complete set of parameters 
to determine their individual apexes.  We computed these individual
apexes and their error ellipses using the technique described by Chupina et al. (2001).

\begin{figure}[!tH]
\vbox{
\centerline{\psfig{figure=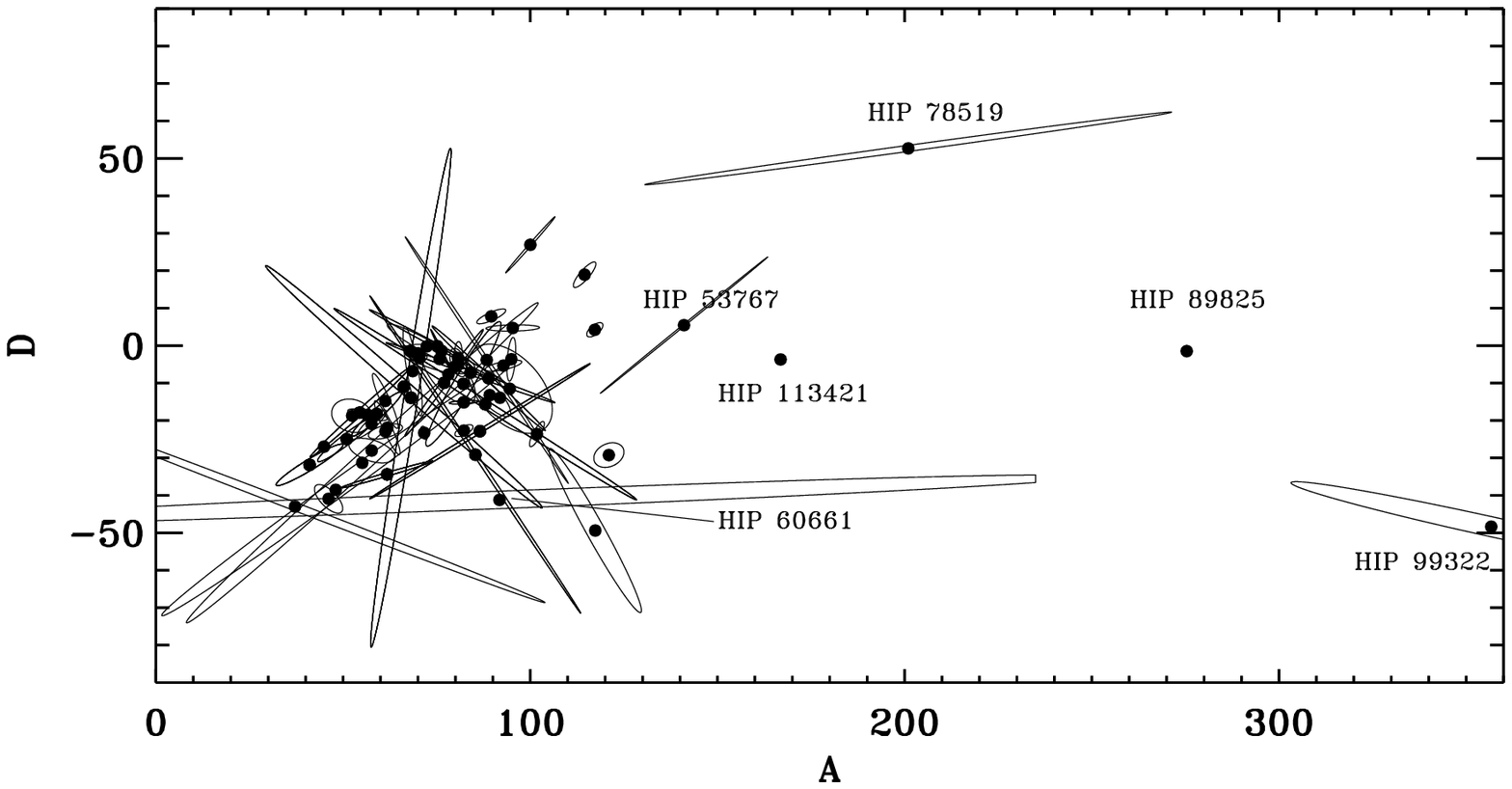,width=12.0cm,angle=0,clip=}}
\vspace{1mm}
\captionb{1}
{
The AD-diagram with the error ellipses for the Castor stream. 
The stars with large errors ellipses are marked by their HIP numbers. 
}
}
\end{figure}

Figure~1 shows the AD-diagram for the Castor stream. 
For most of the stars the error ellipses are small. 
Stars with large error ellipses 
have HIP numbers 29067, 33451, 60661, 72622, 113263, 116132. 
We did not use these stars in our apex determination and excluded them from our list of
stream members. 
The large errors of the parameters of these stars distort the inferred apex, 
and  they are located far from the average apex position in the AD-diagram. 
HIP~60661 is a runaway star (Lopez-Santiago et al. 2010) and its 
membership in the Castor group is not yet clear.

The concentration of points in the central part of Figure~1 is not the 
core of the stream and neither the scattered periphery should be viewed
as its corona. 
The scatter of points is due to the presence of multiple stars whose 
velocities are distorted by the orbital motion. 
Here we do not analyze the features that are apparent 
in the diagram, such as, e.g.,  nonuniform density distribution and 
different orientations and sizes of the axes of the  error ellipses.

We determine the Castor apex by averaging the individual apex values by applying 
an iterative $3\sigma$ clipping: we first compute the mean apex averaged 
over all stars, then discard the stars lying beyond $3\sigma$ and repeat the procedure.
After four iterations we find $A=79.33^\circ$, $D=-15.06^\circ$ based on the data for
65 stars, which are assigned the highest membership probability in our list. 
The apex position is shown in Figure~1 and we discuss it in in Section~5.

\sectionb{4}{COMPARISON OF THE AD-DIAGRAMS FOR THE CASTOR AND UMA STREAMS}

To compare the kinematics of the two streams, we 
plot in Fig.~2 the AD diagram including the data for both the 
Castor and UMa moving groups adopting the data for the latter from 
our earlier paper (Chupina et al. 2001). 
In this AD diagram the UMa stream is represented by single stars exclusively. 
As already noted, this stream includes three groups, 
which are marked in Fig.~2.

\begin{figure}[!tH]
\vbox{
\centerline{\psfig{figure=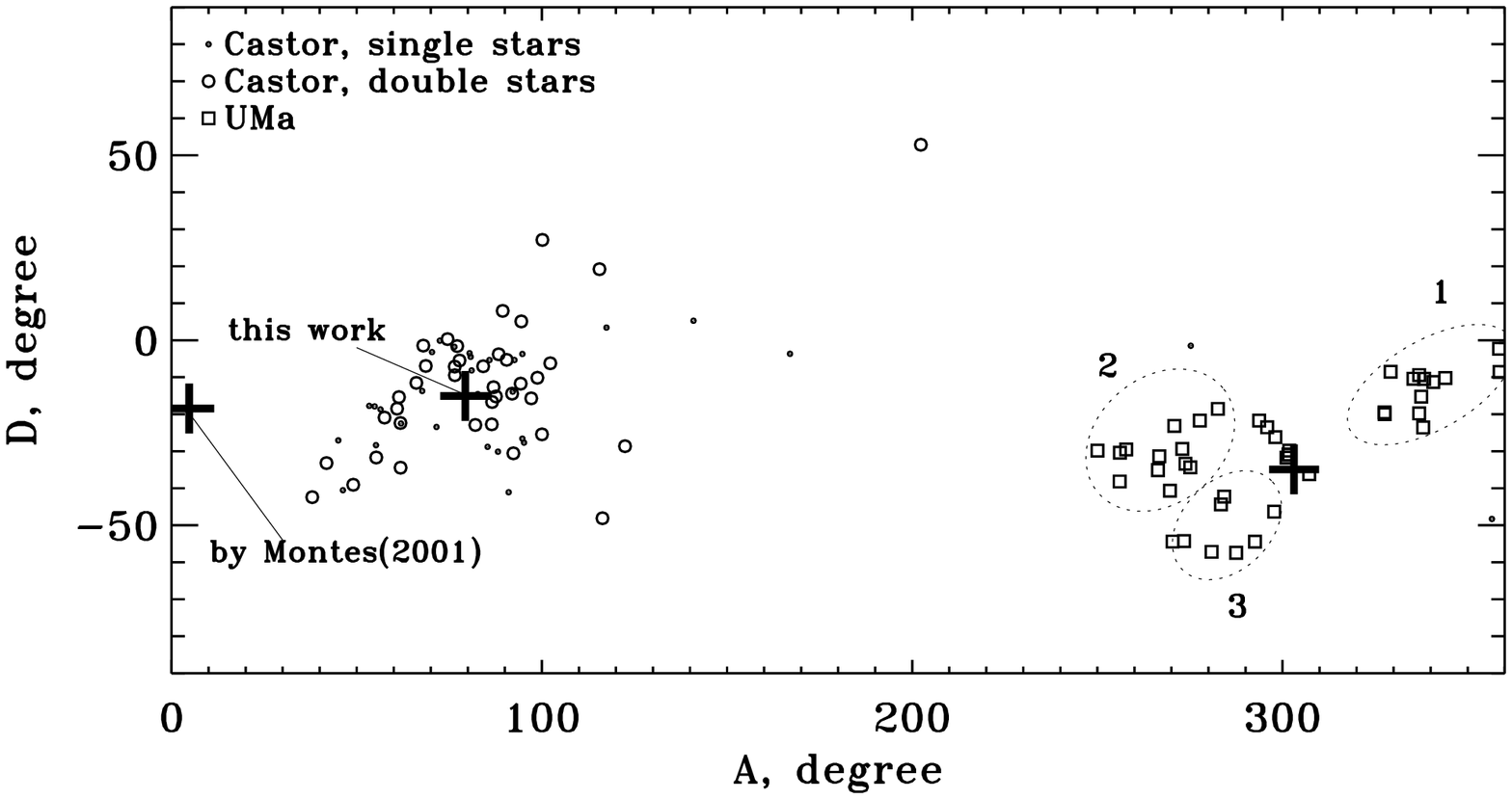,width=12.0cm,angle=0,clip=}}
\vspace{1mm}
\captionb{2}
{
The AD-diagram for the Castor (the small open squares) and UMa (the open circles)
streams. 
The big crosses indicate the stream apex positions. 
}
}
\end{figure}

Why is the dispersion of points for the Castor stream  so large? 
This is because this stream contains many  multiple systems whose 
individual apex positions may be incorrect.  
The angular distance between the UMa and Castor apexes
in Fig.~2 is about $225^\circ$, implying that the two streams
move in opposite directions in space.

\sectionb{5}{STREAMS IN SPACE}

Figure~3 shows the distribution of stars projected onto the Galactic plane
in the Cartesian coordinate system with the origin at the Sun. 
The X-axis points toward the Galactic center; the
Y-axis,  in the direction of Galactic rotation, and the 
Z-axis (not shown here), toward the North Galactic Pole. 
As far as the Z-coordinate is concerned, the streams studied
are located practically in the Galactic midplane 
at the distances of 10.6~pc and 2.4~pc for UMa and Castor, respectively.

\begin{figure}[!tH]
\vbox{
\centerline{\psfig{figure=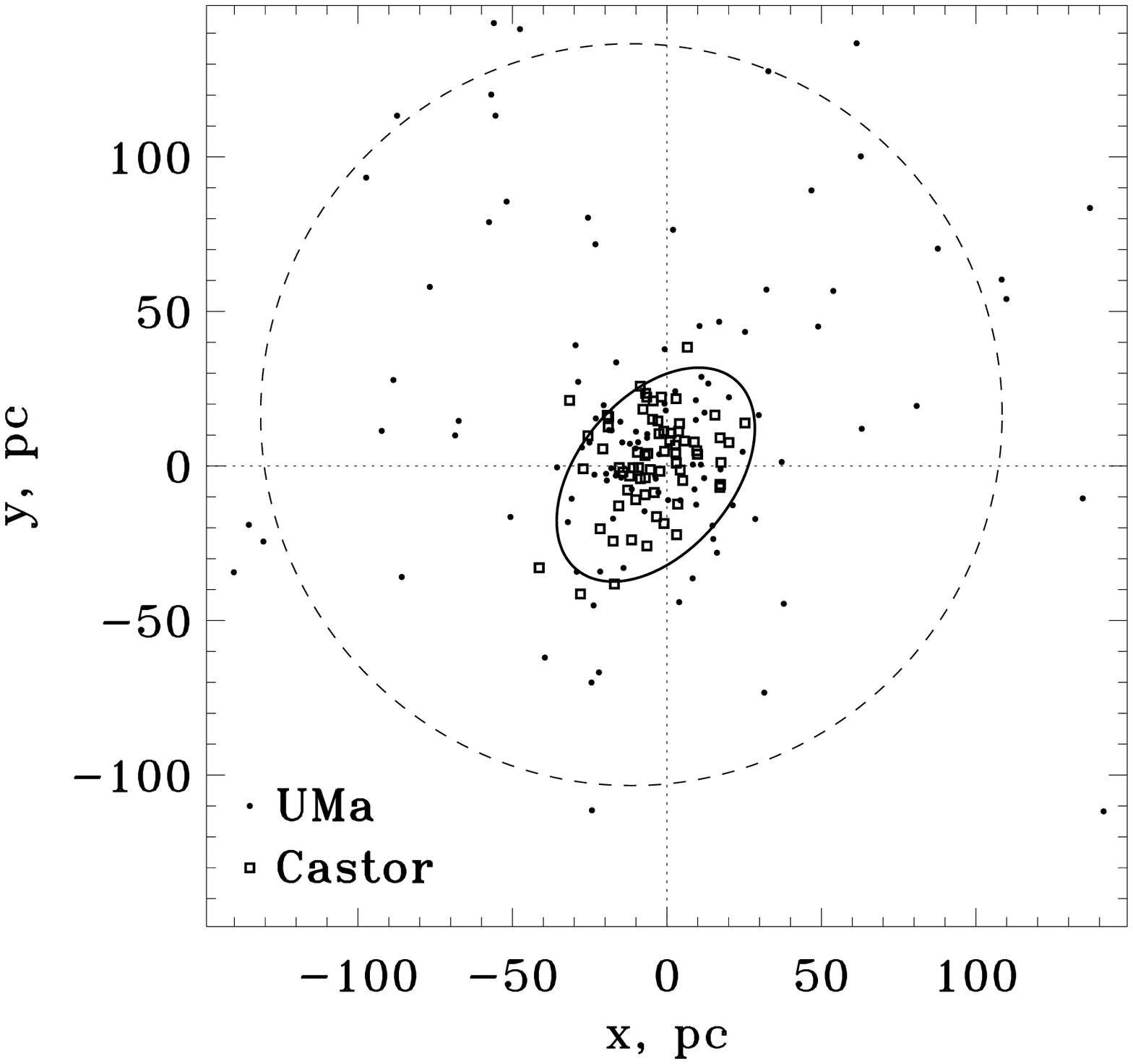,width=8.0cm,angle=0,clip=}}
\vspace{1mm}
\captionb{3}
{ Distribution of stars of the Castor and UMa streams projected onto the XY plane 
(Galactic plane). The Castor and UMa stream stars are shown by the points
and squares, respectively. 
The areas occupied by the UMa and Castor's streams are outlined by the dashed 
and solid lines, respectively.
}
}
\end{figure}

The entire Castor stream is located inside the central part of UMa. 
It has the ellipse-like outlines with the semi-major located almost along
the diagonal of the XY-box. 
The two streams have almost equal space velocities: 
20.1~km/s for UMa and 20.3~km/s for Castor. 

We estimated the above space velocities from the available data radial velocities, 
parallaxes, and proper motions. 
These space velocities can be useful for understanding the physical nature
of the streams studied.

\sectionb{6}{DISCUSSION AND CONCLUSIONS}

Stars of two streams closest to the Sun are considered in phase space.  
These streams are the interesting phenomenon in a Galaxy disk. 
Castor and UMa moving groups interpenetrate each other in space 
and they are moving in opposite directions. 

Can the Castor stream be viewed as a  kinematic groups within the UMa stream? 
Their space velocities differ very much, by more than 40~km/s. 
Obviously, these star systems are not gravitationally bound and will diverge. 
These groups must originate from different parts of the disk. 
The UMa stream and the whole Sirius supercluster move 
in the direction of the antiapex of most of Galactic clusters, whereas 
the position of the Castor stream in the velocity space is close to that of clusters.

Structures like the UMa and Castor streams are not uncommon. 
There are systems of clusters in phase space, e.g.,
the group of clusters in the Orion Sword (Bouy et al. 2014). 
In the same area Group~189 was discovered,
which can belong to the corona of NGC~1977. 
Famaey et al. (2008) detected four structures 
which correspond to Hercules's flow, Pleaydes, Hyades, and Sirius group.

\begin{table}[!t]
\begin{center}
\vbox{\footnotesize\tabcolsep=3pt
\parbox[c]{124mm}{\baselineskip=10pt
{\smallbf\ \ Table 1.}{\small\
Comparison of the Castor apex position determinations \lstrut}}
\begin{tabular}{l|cc}
\hline
author&A, degree&D, degree\\
\hline
This work&79.33&-15.06\\
Montes et al. (2001)&4.75&-18.44\\
\hline
\end{tabular}
}
\end{center}
\vskip-3mm
\end{table}

We determined the position of  the Castor stream apex. 
Montes et al. (2001) report a very different
apex position (see Table~1). This very large discrepancy has to be solved.
It may be 
due to the fact that the above study was based on the data for late-type stars. 
Individual apexes in the AD-diagram deviate strongly from the average value 
with the scatter equal to
$\sigma_A=18.92^\circ$ and $\sigma_D=14.43^\circ$. 
We inferred this Castor apex estimate taking into account both these large scatter values
and parameter errors.

\References

\refb Agekyan, T.A., Orlov, V.V. 1984, AZh, 61, 60
\refb Anosova, J.P., Orlov, V.V. 1991, A\&A,  252, 123
\refb Barrado y Navascues, D. 1998, A\&A, 339, 831  
\refb Bouy, H., Alves, J., Bertin, E., et al. 2014, A\&A, 564, 29
\refb Caballero, J. A. 2010, A\&A, 514, 98
\refb Chupina, N.V., Reva, V.G., and Vereshchagin, S.V. 2001, A\&A, 371, 115  
\refb Famaey, B., Siebert, A., Jorissen, A. 2008, A\&A, 483, 453
\refb Lopez-Santiago, J., Montes, D., Galvez-Ortiz, M.C. et al. 2010, A\&A, 514, 97.
\refb Mamajek, Eric E., Kenworthy, Matthew A., Hinz, Philip M., et al. 2010, Astron.~J, 139, 919
\refb Montes, D., Lopez-Santiago, J., Galvez, M. C., et al. 2001, MNRAS, 328, 45
\refb The Hipparcos and Tycho Catalogues, ESA SP-1200 (ESA, 1997)
\refb Shkolnik, Evgenya L., Anglada-Escude, Guillem, Liu, Michael C., et al. 2012, ApJ, 758, 56
\refb SIMBAD database, Strasbourg, France\\
      http://simbad.u-strasbg.fr/simbad/
\refb Zhuchkov, R.Ya., Orlov, V.V., Rubinov, A.V. 2006, ARep, 50, 62

\end{document}